\documentclass[11pt,onecolumn,oneside]{osajnl}
\journal{jocn} % Choose journal (ao, aop, josaa, josab, ol, optica, pr)

\title{On the Intercept Probability and Secure Outage Analysis of Mixed ($\alpha-\kappa-\mu$)-shadowed and M\'alaga Turbulent Model}

\author[1]{Noor Ahmad Sarker}
\author[1]{A. S. M. Badrudduza}
\author[2]{S. M. Riazul Islam}
\author[3]{Sheikh Habibul Islam}
\author[4]{Milton Kumar Kundu}
\author[5]{Imran Shafique Ansari}
\author[6]{Kyung-Sup Kwak}

\affil[1]{Department of Electronics \& Telecommunication Engineering, Rajshahi University of Engineering \& Technology (RUET), Rajshahi-6204, Bangladesh}
\affil[2]{Department of Computer Science and Engineering, Sejong University, Seoul 05006, South Korea}
\affil[3]{Department of Electrical \& Electronic Engineering,RUET, Rajshahi-6204, Bangladesh}
\affil[4]{Department of Electrical \& Computer Engineering, RUET, Rajshahi-6204, Bangladesh}
\affil[5]{James Watt School of Engineering, University of Glasgow, Glasgow G12 8QQ, United Kingdom}
\affil[6]{School of Information and Communication Engineering, Inha University, Incheon 22212, South Korea}

\begin{abstract}
This work deals with the secrecy performance analysis of a dual-hop RF-FSO DF relaying network composed of a source, a relay, a destination, and an eavesdropper. We assume the eavesdropper is located close to the destination and overhears the relay’s transmitted optical signal. The RF and FSO links undergo ($\alpha-\kappa-\mu$)-shadowed fading and unified M\'alaga turbulence with pointing error. The secrecy performance of the mixed system is studied by deriving closed-form analytical expressions of secure outage probability (SOP), strictly positive secrecy capacity (SPSC), and intercept probability (IP). Besides, we also derive the asymptotic SOP, SPSC, and IP upon utilizing the unfolding of Meijer’s $G$ function where the electrical SNR of the FSO link tends to infinity. Finally, the Monte-Carlo simulation is performed to corroborate the analytical expressions. Our results illustrate that fading, shadowing, detection techniques (i.e. heterodyne detection (HD) and intensity modulation and direct detection (IM/DD)), atmospheric turbulence, and pointing error significantly affect the secrecy performance. In addition, better performance is obtained exploiting the HD technique at the destination relative to IM/DD technique.

\quad

\end{abstract}

\begin{document}

\maketitle

%%%%%%%%%%%%%%%%%%%%%%%%%%%%%%%%%%%%%%%%%%%%%%%%%%%%%%%%%%%%%%%%%%%%%%%%%%%%%%%%%%%%%%%%%%%%%%%%%%%%
%%%%%%%%%%%%%%%%<SECTION>%%%%%%%%%%%%%%%
\section{Introduction}

%#############<SUB-SECTION>#############
\subsection{Background and Related Works}
Free space optical (FSO) technology has drawn significant attention of the research communities compared to traditional radio frequency (RF) technologies in wireless communication applications due to advantages of high-frequency bandwidth, high speed, high security, large transmission capacity, disaster recovery, fast deployment, unlicensed spectrum, back-haul for wireless cellular networks, solution for the last-mile access problem, fiber backup, and no interference, among many others \cite{makki2016performance}. However, pointing error and atmospheric turbulence highly impact the system performance of FSO schemes \cite{al2001mathematical, farid2007outage, 6952039,7883900} that can be mitigated by utilizing a dual-hop mixed RF-FSO relaying system.

Since the wireless medium is time-varying in nature, recently, researchers are devoting their concentrations to composite fading models that can unify the characteristics of a wide range of classical multipath / generalized fading models, thereby applicable to more practical / real-life scenarios \cite{10754/134733}. $\alpha-\mu$ \cite{yacoub2007alpha}, ($\alpha-\mu$)-shadowed \cite{upaddhyayapproximate}, $\kappa-\mu$ \cite{bhargav2016secrecy}, ($\kappa-\mu$)-shadowed \cite{paris2013statistical}, and $\eta-\mu$ \cite{pena2014performance,6692669} are widely used as generalized models in the literature. To obtain further generalization, authors in \cite{fraidenraich2006alpha} proposed $\alpha-\kappa-\mu$ and $\alpha-\eta-\mu$ distributions that were further generalized by $\alpha-\kappa-\eta-\mu$ model \cite{da2017product}. It can be noted that the authors considered randomly fluctuated dominant specular components, non-linearity of the propagation medium, non-line-of-sight (NLOS), line-of-sight (LOS) propagation link, etc., criterion for channel modeling. Considering all of those channel effects, ($\alpha-\kappa-\mu$)-shadowed (AKM-shadowed) model was formulated in \cite{ramirez2019alpha} that possesses a good mathematical tractability and offers a natural generalization to all the aforementioned channel models.

In recent years, researchers have carried out a mesmerizing amount of works on FSO communication systems \cite{malik2015free, willebrand2001fiber, 6966082, sharma2013high, sahbudin2013performance, 7881143}. The analysis of the system performance considering the FSO scheme was first performed in \cite{zhu2002free} focusing on the impact of turbulence-induced fading. This model was further upgraded with multiple receive and transmit apertures in the existence of both background and shot noises \cite{haas2003capacity}. Data transmission using series and parallel relays in FSO communication scheme was introduced in \cite{safari2008relay}. The authors in \cite{uysal2004error, uysal2006error} performed the error control coding for two different FSO models. A Unique multi-input multi-output (MIMO) model was proposed in \cite{farid2011diversity} with multiple transmitters and receivers considering the effect of fading and pointing error. The adverse effect of boresight pointing error on a FSO link for both intensity modulation / direct detection (IM/DD) and heterodyne detection (HD) techniques was analyzed in \cite{ansari2015ergodic}. The unification of the existing FSO models was done by introducing the M\'alaga turbulence model in \cite{ansari2015performance}.

Recently, mixed RF-FSO systems have been investigated thoroughly to eliminate atmospheric turbulence dependency of the FSO links. In such types of scenarios, long-distance communication is performed over the RF hop whereas shorter distance is accomplished over the FSO hop. The authors in \cite{lee2011performance} studied the performance of amplify-and-forward (AF) fixed gain relaying technique in terms of outage probability (OP) considering Rayleigh-$\Gamma\Gamma$ fading scenario. The performance of a nearly similar model was analyzed \cite{zedini2014performance} for both HD and IM/DD techniques. Authors in \cite{zhao2017performance} investigated the impact of aperture averaging of the FSO link. In \cite{soleimani2015generalized, li2019performance, al2019precise}, authors considered both decode-and-forward (DF) and AF relaying methods for dual-hop RF-FSO network and derived closed-form expressions (CFE) for OP, ergodic capacity (EC), and bit error rate (BER). Similar performance parameters were also investigated \cite{gupta2018performance} where the authors choose different fading models as for both RF and FSO link. The increment in atmospheric temperature causes the thermal expansion in the buildings around us which in turn produces non-zero boresight pointing error. Authors in \cite{odeyemiimpact} modeled a RF-FSO system to analyze the impact of such error.

With the rapid growth of wireless networks, secret information transmission in presence of adversaries is an extremely critical issue. The traditional security methods depend on cryptographic techniques at upper layers of wireless networks that are difficult to utilize. In this perspective, physical layer security (PLS) is the only solution that utilizes the randomness of the propagation channel to enhance the secrecy level \cite{mostafa2015physical, wang2018physical, fang2017stackelberg, ibrahim2015relay}. The effect of imperfect channel state information was considered in \cite{lei2018secrecy} and the performance analysis was carried out in terms of secrecy outage probability (SOP) adopting the fixed gain relaying technique. In \cite{lei2017On_secrecy}, the authors observed that RF hop has a little impact on SOP and average secrecy capacity (ASC) performances relative to FSO hop. The position of eavesdropper was considered close to the destination in \cite{pan2019secrecy, pattanayak2020physical2} where the authors presented the expressions of SOP and strictly positive secrecy capacity (SPSC) with DF relaying system. A passive RF eavesdropping scheme was used in mixed RF-FSO systems over $\Gamma\Gamma$ \cite{lei2017secrecy,sarker2020secrecy}, M\'alaga \cite{islam2020secrecy,pattanayak2020physical, mandira2021secrecy}, exponentiate-Weibull \cite{juel2021secrecy,lei2018secrecy}, etc., scenarios to obtain ASC, SOP, and SPSC. The effects of transmit antenna selection (TAS) scheme over the RF hop in a RF-FSO mixed system was examined by \cite{lei2020secure}. In \cite{islam2021impact, sarker2021effects}, authors compared the performance between RF and FSO eavesdropping over M\'alaga and double generalized Gamma (DGG) models and demonstrated that FSO technology is more secure than RF technology.

%#############<SUB-SECTION>#############
\subsection{Motivation and Contributions}
Based on aforecited literature,it is seen that among the existing PLS works on RF-FSO schemes, RF hop is typically assumed to experience multipath / generalized fading while none of these works considered the impact of shadowing on the RF hop.
%Even a few works considered eavesdropper's placing at the FSO link.
In this work, we consider a mixed RF-FSO dual-hop DF relaying system where the RF and FSO links, respectively, experience AKM-shadowed fading and M\'alaga turbulence fading model included with pointing error. We consider the position of the eavesdropper is very close to the destination and can decipher the transmitted optical signals from the relay. Our main contributions in this work are pointed as follows.

\begin{itemize} 

\item We first realize the probability density functions (PDFs) and cumulative distribution functions (CDFs) of the AKM-Shadowed link and M\'alaga turbulence link for the individual hops of the considered dual-hop system. Since our considered RF and FSO models account for a high form of generality, our obtained results can be ascertained as a generalization of the existing results in \cite{pattanayak2020physical, pan2019secrecy}.

\item To analyze the secrecy performance, we derive the CFEs for the SOP, SPSC, and intercept probability (IP). To obtain more practical insights, the asymptotic expressions for these performance parameters are also provided. Finally, we present Monte-Carlo simulations to verify the accuracy of the CFEs.

\item Capitalizing on the final expressions of the secrecy performance parameters, we observe impacts of fading, shadowing, atmospheric turbulence, pointing error, etc., on the secrecy of the proposed scenario. Additionally, we also present a comparison between the performance of two detection techniques i.e. HD and IM/DD techniques.

\end{itemize}

%#############<SUB-SECTION>#############
\subsection{Organization}
The rest of the paper is arranged in the following manner. The system model is described in Section II including formulations of the fading channels. In Section III, expressions for SOP, SPSC, and IP are derived in both exact and asymptotic forms. Analytical and simulation results are presented in Section IV, and finally, the concluding remarks are provided in Section V.

\section{System Model and Problem Formulation}
%===============<FIGURE>================
\begin{figure}[!h]
\vspace{0mm}
    \centerline{\includegraphics[width=0.8\textwidth,angle =0]{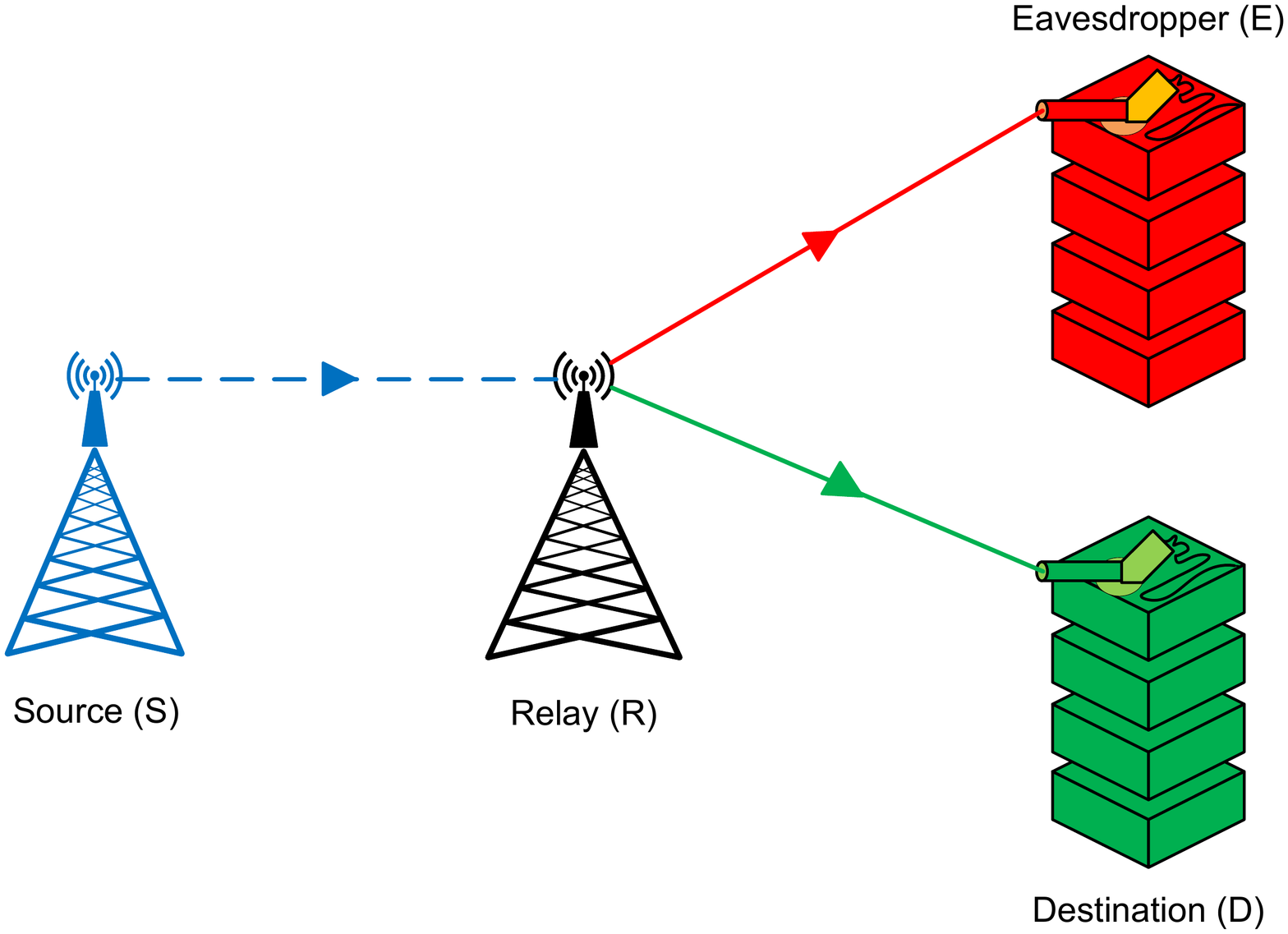}}
    \vspace{0mm }
    \caption{The dual-hop mixed RF-FSO system.}
    \label{f01}
\end{figure}
%=============<END-FIGURE>==============
We consider a combined RF-FSO DF-based relaying system as demonstrated in Fig. \ref{f01}. Here, information is transferred from a stable source $S$ to a destination $D$ via a relay $R$ that works as an intermediate medium between $S$ and $D$. It is considered that an unexpected eavesdropper $E$ tries to hijack the classified information from $R$ that is supposed to reach $D$. Similar to \cite{saber2017secure}, we presume that $E$ is firmly located around $D$. $S$ with a single antenna transmits information to $R$ through an independent and identically distributed (i.i.d.) AKM-shadowed RF fading link where $R$ consists of one receiving antenna and one transmit aperture. After receiving the RF signal, $R$ converts the same to an optical signal and then re-transmits it. Both $R-D$ and $R-E$ links are connected linked via FSO technology experiencing M\'alaga ($\mathcal{M}$) turbulence with pointing error. Here, $D$ and $E$ both contain one receive aperture for receiving the optical signals.

%#############<SUB-SECTION>#############
\subsection{SNRs of Each Link}
As for the considered communication scenario in Fig. \ref{f01}, the instantaneous signal-to-noise ratios (SNRs) denoted by $S-R$, $R-D$, and $R-E$ links are $\gamma_{r}$, $\gamma_{d}$, and $\gamma_{e}$, respectively. These terms take arithmetic forms such as $\gamma_{r}=\phi_{r}\|\beta_{r}\|^{2}$, $\gamma_{d}=\phi_{d}\|\beta_{d}\|^{2}$, and $\gamma_{e}=\phi_{e}\|\beta_{e}\|^{2}$, where $\beta_{r}, \beta_{d}$, and $\beta_{e}$ represent channel gains, and   $\phi_{r}, \phi_{d}$, and $\phi_{e}$ represent average SNRs of the $S-R$, $R-D$, and $R-E$ links, respectively. As relay $R$ employs DF relaying scheme, end-to-end SNRs for both main $S-R-D$ receiver link and eavesdropper $S-R-E$ receiver link are formulated as \cite[Eq.~(5),]{hasna2004performance}
%***************<EQUATION>**************
\begin{subequations}
\begin{align}
&\gamma_{sd}=min\left\{\gamma_{r}, \gamma_{d}\right\},
\\
&\gamma_{se}=min\left\{\gamma_{r}, \gamma_{e}\right\}.
\end{align}
\end{subequations}
%*************<END-EQUATION>************

%#############<SUB-SECTION>#############
\subsection{RF Channel}
As the link between $S$ and $R$ experiences AKM-shadowed distribution, the PDF can be expressed as \cite[Eq.~(4),]{ramirez2019alpha}
%***************<EQUATION>**************
\begin{align}  
\label{a1}
f_{\gamma_{r}}(\gamma)=\mathcal{A}_{1}\gamma^{\frac{\alpha\mu}{2}-1}e^{-\mathcal{A}_{2}\gamma^{\alpha/2}}\,_1F_1(x,\mu;\mathcal{A}_{3}\gamma^{\alpha/2}),
\end{align}
%*************<END-EQUATION>************
where $\mathcal{A}_{1}=\frac{x^{x}\alpha\phi_{r}^{-\alpha\mu/2}}{2d^{\mu}\Gamma(\mu)(\mu\kappa+x)^{x}}$, $\mathcal{A}_{2}=\frac{1}{d\phi_{r}^{\alpha/2}}$, and $\mathcal{A}_{3}=\frac{\mu\kappa}{d(\mu\kappa+x)\phi_{r}^{\alpha/2}}$. The terms $x$, $\mu$, $\alpha$, and $\kappa$ are all real shape parameters with non-negative values i.e. $\alpha$ is a power term indicating non-linearity nature of signal envelop \cite{yacoub2007alpha}, $\kappa$ defines the ratio of total amount of powers between dominant and scattered waves \cite{paris2013statistical}, and $\mu$ and $x$ denote number of cluster and fading severity parameters, respectively, \cite{ramirez2019alpha}. $\Gamma(.)$ represents Gamma function \cite[Eq.~(8.310),]{Calculationbook01}. The function $_1F_1(.)$ is the confluent hypergeometric function as defined in \cite[Eq.~(13.1.2),]{hyperfunction}. The constant term $d$ is described as
%***************<EQUATION>**************
\begin{align}  
\nonumber
d=\biggl[\frac{(\mu\kappa+x)^{x}\Gamma(\mu)}{x^{x}\Gamma(\mu+\frac{2}{\alpha})\,_2F_1(x,\mu+\frac{2}{\alpha};\mu;\frac{\mu\kappa}{\mu\kappa+x})}\biggl]^{\alpha/2}.
\end{align}
%*************<END-EQUATION>************
Here, the function $_2F_1(.)$ is the Gauss hypergeometric function as defined in \cite[Eq.~(15.1.1),]{hyperfunction}. AKM-shadowed is a composite fading model that houses many multipath and generalized fading channel models. Such a model promises to provide more insights over a wide range of variations of channel conditions that is treated as more practical propagation environment by the wireless communication researchers \cite{7145711, 7145973}. Table \ref{t1} lists some familiar RF fading channels that can be obtained as special cases of AKM-shadowed fading channel.
%<<<<<<<<<<<<<<<<<TABLE>>>>>>>>>>>>>>>>>
\begin{table}[!h]
\centering
\caption{Special Cases of AKM-Shadowed Composite Fading Channel \cite[Table I,]{ramirez2019alpha}.}
\scalebox{1.10}{%
\begin{tabular}{|c|c|}
\hline
\multicolumn{1}{|c|}{Channels} & \multicolumn{1}{c|}{AKM-Shadowed Fading Parameters}\\
\hline
\hline
Rayleigh & $\alpha=2$, $\kappa=0$, $\mu=1$
\\
\hline 
Nakagami-$m$ & $\alpha=2$, $\kappa=0$, $\mu=m$
\\
\hline
$\kappa-\mu$ & $\alpha=2$, $\kappa=\kappa$, $\mu=\mu$, $x\rightarrow\infty$
\\
\hline 
$\eta-\mu$ & $\alpha=2$, $\kappa=(1-\eta)/(2\eta)$, $\mu=2\mu$, $x=\mu$
\\
\hline
Weibull & $\alpha=\alpha$, $\kappa=0$, $\mu=1$
\\
\hline
$\alpha-\kappa-\mu$ & $\alpha=\alpha$, $\kappa=\kappa$, $\mu=\mu$, $x\rightarrow\infty$
\\
\hline
\end{tabular}}
\label{t1}
\end{table}
%<<<<<<<<<<<<<<<END-TABLE>>>>>>>>>>>>>>>
Utilizing \cite[Eq.~(9.14.1),]{Calculationbook01}, \eqref{a1} is alternatively expressed as
%***************<EQUATION>**************
\begin{align}
\label{ab1}
f_{\gamma_{r}}(\gamma)=\mathcal{A}_{1}\sum_{i=0}^{\infty}\mathcal{A}_{4}\gamma^{\frac{\alpha(\mu+i)}{2}-1}e^{-\mathcal{A}_{2}\gamma^{\frac{\alpha}{2}}},
\end{align}
%*************<END-EQUATION>************
where $\mathcal{A}_{4}=\frac{x^{i}\mathcal{A}_{3}^{i}}{\mu^{i}i!}$. The CDF of this channel can be found by utilizing
%***************<EQUATION>**************
\begin{align}
\label{a2}
F_{\gamma_{r}}(\gamma)=\int_{0}^{\gamma}f_{\gamma_{r}}(\gamma)d\gamma.
\end{align}
%*************<END-EQUATION>************
Placing \eqref{ab1} into \eqref{a2}, utilizing \cite[eqs.~(3.381.8) and (8.352.6),]{Calculationbook01}, and performing integration, \eqref{a2} is obtained as
%***************<EQUATION>**************
\begin{align}  
\label{a3}
F_{\gamma_{r}}(\gamma)=1-\frac{2\mathcal{A}_{1}}{\alpha}\sum_{i=0}^{\infty}\sum_{j=0}^{\mu+i-1}\mathcal{A}_{5}\gamma^{\frac{\alpha j}{2}}e^{-\mathcal{A}_{2}\gamma^{\frac{\alpha}{2}}},
\end{align}
%*************<END-EQUATION>************
where $\mathcal{A}_{5}=\frac{\mathcal{A}_{4}\Gamma(\mu+i)}{j!\mathcal{A}_{2}^{\mu+i-j}}$.

%#############<SUB-SECTION>#############
\subsection{FSO Channel}
Now, PDF of FSO link, formulated by unified $\mathcal{M}$ turbulence model, is expressed as \cite[Eq.~(9),]{ansari2015performance}
%***************<EQUATION>**************
\begin{align}
\label{a4}
f_{\gamma_{m}}(\gamma)=\frac{\mathcal{Z}_{1}}{\gamma}\sum_{q_{m}=0}^{b}h_{q_{m}}G_{1,3}^{3,0}\left[\mathcal{Z}_{2}\left(\frac{\gamma}{U_{m}}\right)^{\frac{1}{s_{m}}}\biggl|
\begin{array}{c}
\epsilon^2+1\\
\epsilon^2,a,q_{m}\\
\end{array}
\right],
\end{align}
%*************<END-EQUATION>************
where $m\in\left\{d,e\right\}$ correspond to $R-D$ and $R-E$ links, respectively,
%***************<EQUATION>**************
\begin{align}  
\nonumber
\mathcal{Z}_{1}&=\frac{2^{1-s_{m}}\epsilon^{2}a^{a/2}}{r_{m}^{1+\frac{a}{2}}\Gamma(a)}\biggl(\frac{r_{m}b}{r_{m}b+\zeta_{t_{m}}}\biggl)^{b+\frac{a}{2}},
\\
\nonumber
\mathcal{Z}_{2}&=\frac{\epsilon^{2}ab(r_{m}+\zeta_{t_{m}})}{(\epsilon^{2}+1)(r_{m}b+\zeta_{t_{m}})},
\\
\nonumber
h_{q_{m}}&=j_{q_{m}}\biggl(\frac{ab}{r_{m}b+\zeta_{t_{m}}}\biggl)^{-\frac{a+q_{m}}{2}},
\\
\nonumber
j_{q_{m}}&=\binom{b-1}{q_{m}-1}\frac{(r_{m}b+\zeta_{t_{m}})^{1-\frac{q_{m}}{2}}}{(q_{m}-1)!}\biggl(\frac{\zeta_{t_{m}}}{r_{m}}\biggl)^{q_{m}-1}\biggl(\frac{a}{b}\biggl)^{\frac{q_{m}}{2}},
\end{align}
%*************<END-EQUATION>************
$\mathcal{Z}\in\left\{\mathcal{B},\mathcal{C}\right\}$ correspond to $R-D$ and $R-E$ links, respectively, $a$ and $b$ both are related to the turbulence conditions in atmosphere with $a$ being firmly related to the effective number that is followed by the large-scale cells and $b$ being related to the scattering process \cite{zedini2016onthe}, $\epsilon$ is identified as a ratio between the tantamount beam radius signal beam and the standard deviation (jitter) due to pointing error misalignment, $s_{m}$ represents detection type at the receiver (i.e. $s_{m}=1$ for HD technique and $s_{m}=2$ for IM/DD technique), $U_{m}$ represents the electrical SNR of FSO link that is expressed dependent on the detection technique $s_{m}$ such that $U_{1}=\phi_{m}$ for HD technique and $U_{2}=\frac{a\epsilon^{2}(\epsilon^{2}+1)^{-2}(\epsilon^{2}+2)(r_{m}+\zeta_{t_{m}})}{(a+1)[2r_{m}(r_{m}+2\zeta_{t_{m}})+\zeta^{2}_{t_{m}}(1+1/b)]}\phi_{m}$ for IM/DD technique \cite{ansari2015performance}, $r_{m}$ represents the average power of scattering components that is received by the off-axis eddies subjected to the FSO link \cite{jurado2011unifying}, $\zeta_{t_{m}}=\zeta_{m}+2h_{0m}\varrho+\sqrt{2h_{0m}\varrho\zeta_{m}}cos(\theta_{x_{m}}-\theta_{y_{m}})$ i.e. average power subjected to the coherent contributions in the FSO link, $\zeta_{m}=2h_{0m}(1-\varrho)$ represents the average power of LOS component, $2h_{0m}$ is the average power of all scattered components, $\varrho$ denotes the total amount of scattering coupled power placed at LOS component with the limit $0\leq\varrho\leq1$, $\theta_{x_{m}}$ and $\theta_{y_{m}}$ both are the LOS deterministic phases \cite{jurado2011unifying}, and G[.] symbolizes the Meijer's $G$ function as defined in \cite[Eq.~(9.301),]{Calculationbook01}. M\'alaga distribution is one of the most popular FSO fading models among optical wireless communication researchers due to its outstanding generic characteristics. Table \ref{t2} lists some classical FSO fading channels that can be obtained from the unified $\mathcal{M}$ turbulence model via tuning some of its parameters.
%<<<<<<<<<<<<<<<<<TABLE>>>>>>>>>>>>>>>>>
\begin{table}[!h]
\centering
\caption{Some Special Cases of $\mathcal{M}$ Turbulence Fading Channel \cite[Table 1,]{jurado2011unifying}}
\scalebox{1.10}{%
\begin{tabular}{|c|c|}
\hline
\multicolumn{1}{|c|}{Channels} & \multicolumn{1}{c|}{$\mathcal{M}$ Turbulence Fading Parameters} \\ 
\hline
\hline
$\Gamma\Gamma$  & $\varrho=1$, $r_{m}=0$, $\zeta_{t_{m}}=1$
\\
\hline 
Rice-Nakagami          & $\varrho=0$
\\
\hline
Lognormal        & $\varrho=0$, $r_{m}\rightarrow0$
\\
\hline 
$K$ distribution & $\varrho=0$, $b=1$
\\
\hline
\end{tabular}}
\label{t2}
\end{table}
%<<<<<<<<<<<<<<<END-TABLE>>>>>>>>>>>>>>>

\noindent
The CDF for this link is expressed as
\cite[Eq.~(11),]{ansari2015performance}
%***************<EQUATION>**************
\begin{align}  
\label{a5}
F_{\gamma_{m}}(\gamma)=\mathcal{Z}_{3}\sum_{q_{m}=0}^{b}w_{q_{m}}G_{s_{m}+1,3s_{m}+1}^{3s_{m},1}\left[\frac{\mathcal{Z}_{4}}{U_{m}}\gamma\biggl|
\begin{array}{c}
1,l_{m_{1}}\\
l_{m_{2}},0\\
\end{array}
\right],
\end{align}
%*************<END-EQUATION>************
where $\mathcal{Z}_{3}=\frac{\mathcal{Z}_{1}}{(2\pi)^{s_{m}-1}}$, $w_{q_{m}}=h_{q_{m}}s_{m}^{a+q_{m}-1}$, $\mathcal{Z}_{4}=\frac{\mathcal{Z}^{s_{m}}_{2}}{s_{m}^{2s_{m}}}$. Both $l_{m_{1}}$ and $l_{m_{2}}$ are series, described as $l_{m_{1}}=\left\{\frac{\epsilon^{2}+1}{s_{m}},...,\frac{\epsilon^{2}+s_{m}}{s_{m}}\right\}$ containing $s_{m}$ terms and $l_{m_{2}}=\left\{\frac{\epsilon^{2}}{s_{m}},...,\frac{\epsilon^{2}+s_{m}-1}{s_{m}},\frac{a}{s_{m}},...,\frac{a+s_{m}-1}{s_{m}},\frac{q_{m}}{s_{m}},...,\frac{q_{m}+s_{m}-1}{s_{m}}\right\}$ containing $3s_{m}$ terms.

%#############<SUB-SECTION>#############
\subsection{Secrecy Capacity}
For secured transmission between $S$ to $D$ via intermediate relay $R$, we have to find out the secrecy rate of the system wherein confidential and secrete information can be transmitted by dumping the unwanted effects of the eavesdropper. For the considered dual-hop system in Fig. \ref{f01}, secrecy capacity (SC) must be defined for both hops (i.e. $S-R$ and $R-D$). The network in Fig. \ref{f01} demonstrates first hop is independent of the effects of eavesdropper, so instantaneous SC for the RF hop is defined as
%***************<EQUATION>**************
\begin{align} 
\mathcal{T}_{SR}=\frac{1}{2}\log_{2}(1+\gamma_{r}).
\end{align}
%*************<END-EQUATION>************
For main FSO link that is largely affected by the eavesdropper, instantaneous SC for FSO hop is defined as
%***************<EQUATION>**************
\begin{align} 
\mathcal{T}_{RD}=\biggl[\frac{1}{2}\biggl\{\log_{2}(1+\gamma_{d})-\log_{2}(1+\gamma_{e})\biggl\}\biggl]^{+},
\end{align}
%*************<END-EQUATION>************
where $[z]^{+}=max\left\{z,0\right\}$. For DF-based relaying network, the system considered in Fig. \ref{f01} can be described as a series system where such dual-hop network will usually be dominated by worst hop and the instantaneous SC is expressed as \cite[Eq.~(13),]{08610004}
%***************<EQUATION>**************
\begin{align}
\label{a8}
\mathcal{T}_{SD}=min(\mathcal{T}_{SR},\mathcal{T}_{RD}).
\end{align}
%*************<END-EQUATION>************

%%%%%%%%%%%%%%%%<SECTION>%%%%%%%%%%%%%%%
\section{Performance Analysis}
In this section, we derive closed-form expressions for SOP, SPSC, and IP in both exact and asymptotic forms.

%#############<SUB-SECTION>#############
\subsection{Secrecy Outage Probability}
SOP is an important and crucial performance metric for secrecy measurement in wireless systems. It is basically a parameter that indicates the probability of the instantaneous SC falling below the target SC ($\mathcal{T}_{c}$). For the proposed RF-FSO relaying system, SOP can be defined as \cite{badrudduza2020enhancing}
%***************<EQUATION>**************
\begin{align}
\label{a9}
SOP=\Pr\left\{ \mathcal{T}_{SD} < \mathcal{T}_{c} \right\}.
\end{align}
%*************<END-EQUATION>************
We can rewrite \eqref{a9} by using \eqref{a8} as
%***************<EQUATION>**************
\begin{align} 
\nonumber
SOP&=\Pr\left\{\min(\mathcal{T}_{SR},\mathcal{T}_{RD})<\mathcal{T}_{c}\right\}
\\
\nonumber
&=1-\Pr\left\{\min(\mathcal{T}_{SR},\mathcal{T}_{RD})\geq\mathcal{T}_{c}\right\}
\\
\label{a10}
&=1-\Pr\left\{\mathcal{T}_{SR}\geq\mathcal{T}_{c}\right\}\Pr\left\{\mathcal{T}_{RD}\geq\mathcal{T}_{c}\right\}.
\end{align}
%*************<END-EQUATION>************
Substituting \eqref{a3}-\eqref{a5} into \eqref{a10}, we have
%***************<EQUATION>**************
\begin{align}
\label{a11}
SOP&=\int_{0}^{\infty}F_{\gamma_{d}}(\varphi\gamma+\varphi-1)f_{\gamma_{e}}(\gamma)(1-F_{\gamma_{r}}(\varphi-1))+F_{\gamma_{r}}(\varphi-1)d\gamma,
\end{align}
%*************<END-EQUATION>************
where $\varphi=2^{2\mathcal{T}_{c}}$. Due to mathematical complexities, we derive the SOP at lower bound. Letting the condition $\gamma_{e}\rightarrow\infty$, the lower bound of SOP can be evaluated from \eqref{a11} as \cite{ibrahim2021enhancing}
%***************<EQUATION>**************
\begin{align}
\label{a12}
SOP\geq SOP_{L}&\cong\int_{0}^{\infty}F_{\gamma_{d}}(\varphi\gamma)f_{\gamma_{e}}(\gamma)(1-F_{\gamma_{r}}(\varphi-1))+F_{\gamma_{r}}(\varphi-1)d\gamma.
\end{align}
%*************<END-EQUATION>************
Plugging \eqref{a3}-\eqref{a5} into \eqref{a12} and integrating utilizing \cite[Eq.~(2.24.1.1),]{Calculationbook02} by means of some mathematical simplifications, actual expression of the lower bound of SOP is obtained as
%***************<EQUATION>**************
\begin{align}
\label{a13}
SOP_{L}=&1-\sum_{i=0}^{\infty}\sum_{j=0}^{\mu+i-1}\Re\biggl(1-\underbrace{\mathcal{B}_{3}\mathcal{C}_{3}\sum_{q_{e}=1}^{b}\sum_{q_{d}=1}^{b}w_{q_{e}}w_{q_{d}}}_{\triangleq\sum_{q_{d,e}}}G_{s_{\mathcal{M}}+1,s_{\mathcal{E}}+1}^{3s_{e}+1,3s_{d}}\left[\frac{\mathcal{C}_{4}U_{d}}{\mathcal{B}_{4}\varphi U_{e}}\biggl|
\begin{array}{c}
1-l_{d_{2}},1,l_{e_{1}} \\
l_{e_{2}},0,1-l_{d_{1}} \\
\end{array}
\right]\biggl),
\end{align}
%*************<END-EQUATION>************
where $s_{\mathcal{M}}=s_{e}+3s_{d}$, $s_{\mathcal{E}}=3s_{e}+s_{d}$, and $\Re={2\mathcal{A}_{1}\mathcal{A}_{5}(\varphi-1)^{\frac{\alpha}{2}}{exp(-{\mathcal{A}_{2}(\varphi-1)^{\frac{\alpha}{2}}})}}/{\alpha}$. It is noted the expression in \eqref{a13} can be reduced to \cite[Eq.~(19),]{pan2019secrecy} for Rayleigh-$\Gamma\Gamma$ considering the conditions ($\alpha=2$, $\kappa=r_{d}=r_{e}=0$, $\mu=\varrho=\zeta_{t_{d}}=\zeta_{t_{e}}=1$) and to (Nakagami-$m$)-M\'alaga of \cite[Eq.~(35),]{pattanayak2020physical} considering the conditions ($\alpha=\mu=2$, $\kappa=0$).

\vspace{2mm}
%***************<HEADING>***************
\noindent
\textbf{Asymptotic Expression:}

To get better analytical and tractable understanding on secrecy performance, we derive asymptotic expressions of our secrecy performance metrics by considering the condition $U_{m}\rightarrow\infty$. Applying the formula given in \cite[Eq.~(29),]{pattanayak2020physical} and performing some mathematical manipulations on the Meijer's $G$ function in \eqref{a13}, the asymptotic expression of lower bound SOP is obtained as
%***************<EQUATION>**************
\begin{align}
\label{a14}
SOP_{\infty}=&1-\sum_{i=0}^{\infty}\sum_{j=0}^{\mu+i-1}\Re\left[1-{\sum_{q_{d,e}}}\sum_{p=1}^{3s_{d}}\frac{\prod_{l=1,l\neq p}^{3s_{d}}\Gamma(L_{1,p}-L_{1,l})\prod_{l=1}^{3s_{e}+1}\Gamma(1+L_{2,l}-L_{1,p})}{\prod_{l=3s_{d}+1}^{s_{\mathcal{M}}}\Gamma(1+L_{1,l}-L_{1,p})\prod_{l=3s_{e}+2}^{s_{\mathcal{E}}}\Gamma(L_{1,p}-L_{2,l})}\frac{}{}\biggl(\frac{\mathcal{C}_{4}U_{d}}{\mathcal{B}_{4}\varphi U_{e}}\biggl)^{{L_{1,p}}-1}\right],
\end{align}
%*************<END-EQUATION>************
where $L_1=(1-l_{d_{2}},1,l_{e_{1}})$ and $L_2=(l_{e_{2}},0,1-l_{d_{1}})$. The asymptotic expression in \cite[Eq.~(36),]{pattanayak2020physical} can be obtained from \eqref{a14} with $\alpha=\mu=2$ and $\kappa=0$.

%#############<SUB-SECTION>#############
\subsection{Strictly Positive Secrecy Capacity}
For ensuring a secure communication, SPSC is one of the fundamental parameters that is used to place importance to the existence of the SC. According to \cite{liu2013probability}, the probability of SPSC can be defined as
%***************<EQUATION>**************
\begin{align}
\nonumber
SPSC&=\Pr\left\{min(\mathcal{T}_{SR},\mathcal{T}_{RD})>0\right\}
\\
\label{a15}
&=\Pr\left\{\mathcal{T}_{SR}>0\right\}\Pr\left\{\mathcal{T}_{RD}>0\right\}.
\end{align}
%*************<END-EQUATION>************
The two probability terms defined in \eqref{a15} can be evaluated as
%***************<EQUATION>**************
\begin{align}
\nonumber
\Pr\left\{\mathcal{T}_{SR}>0\right\}&=\Pr\left\{\frac{1}{2}\log_{2}(1+\gamma_{r})>0\right\}
\\
\nonumber
&=\Pr\left\{\gamma_{r}>0\right\}
\\
\label{a16}
&=1,
\end{align}
%*************<END-EQUATION>************
and
%***************<EQUATION>**************
\begin{align}
\nonumber
\Pr\left\{\mathcal{T}_{RD}>0\right\}&=\Pr\left\{\frac{1}{2}\biggl\{\log_{2}(1+\gamma_{d})-\log_{2}(1+\gamma_{e})\biggl\}>0\right\}
\\
\nonumber
&=\Pr\left\{\gamma_{d}>\gamma_{e}\right\}
\\
\label{a17}
&=1-\int_{0}^{\infty}F_{\gamma_{d}}(\gamma)f_{\gamma_{e}}(\gamma)d\gamma.
\end{align}
%*************<END-EQUATION>************
Plugging \eqref{a16} and \eqref{a17} into \eqref{a15}, we get
%***************<EQUATION>**************
\begin{align}
\label{a18}
SPSC=1-\int_{0}^{\infty}F_{\gamma_{d}}(\gamma)f_{\gamma_{e}}(\gamma)d\gamma.
\end{align}
%*************<END-EQUATION>************
Placing the values of \eqref{a4} and \eqref{a5} into \eqref{a18}, performing integration utilizing \cite[Eq.~(2.24.1.1),]{Calculationbook02}, and employing mathematical simplifications, the exact form of \eqref{a18} is evaluated to
%***************<EQUATION>**************
\begin{align}
\label{a19}
SPSC=1-{\sum_{q_{d,e}}}G_{s_{\mathcal{M}}+1,s_{\mathcal{E}}+1}^{3s_{e}+1,3s_{d}}\left[\frac{\mathcal{C}_{4}U_{d}}{\mathcal{B}_{4}U_{e}}\biggl|
\begin{array}{c}
1-l_{d_{2}},1,l_{e_{1}}\\
l_{e_{2}},0,1-l_{d_{1}}\\
\end{array}
\right].
\end{align}
%*************<END-EQUATION>************
The expression of SPSC as given in \eqref{a19} can be reduced to the Rayleigh-$\Gamma\Gamma$ scenario \cite[Eq.~(23),]{pan2019secrecy} with $\alpha=2$, $\kappa=r_{d}=r_{e}=0$, $\mu=\varrho=\zeta_{t_{d}}=\zeta_{t_{e}}=1$.

\vspace{2mm}
%***************<HEADING>***************
\noindent
\textbf{Asymptotic Expression:}

Utilizing similar process to \eqref{a14}, the asymptotic expression of SPSC in \eqref{a19} is derived as
%***************<EQUATION>**************
\begin{align}
\label{a20}
SPSC_{\infty}=1-{\sum_{q_{d,e}}}\sum_{p=1}^{3s_{d}}\frac{\prod_{l=1,l\neq p}^{3s_{d}}\Gamma(L_{1,p}-L_{1,l})\prod_{l=1}^{3s_{e}+1}\Gamma(1+L_{2,l}-L_{1,p})}{\prod_{l=3s_{d}+1}^{s_{\mathcal{M}}}\Gamma(1+L_{1,l}-L_{1,p})\prod_{l=3s_{e}+2}^{s_{\mathcal{E}}}\Gamma(L_{1,p}-L_{2,l})}\biggl(\frac{\mathcal{C}_{4}U_{d}}{\mathcal{B}_{4}U_{e}}\biggl)^{{L_{1,p}}-1}.
\end{align}
%*************<END-EQUATION>************

%#############<SUB-SECTION>#############
\subsection{Intercept probability}

The probability at which the eavesdropper succeeds in intercepting the data upheld at the actual receiving device is addressed as intercept probability (IP). It basically indicates the probability that SC is less than zero. For the proposed communication scenario, IP can be mathematically defined as \cite[Eq.~(31),]{pandey2020physical}
%***************<EQUATION>**************
\begin{align}
\nonumber
IP&=\Pr\left\{\mathcal{T}_{RD}<0\right\}
\\
\nonumber
&=\Pr\left\{\gamma_{d}<\gamma_{e}\right\}
\\
\label{a21}
&=\int_{0}^{\infty}F_{\gamma_{d}}(\gamma)f_{\gamma_{e}}(\gamma)d\gamma
\end{align}
%*************<END-EQUATION>************
Plugging \eqref{a4} and \eqref{a5} into \eqref{a21}, performing integration utilizing \cite[Eq.~(2.24.1.1),]{Calculationbook02}, and simplifying the expression, IP is evaluated to
%***************<EQUATION>**************
\begin{align}
\label{a22}
IP={\sum_{q_{d,e}}}G_{s_{\mathcal{M}}+1,s_{\mathcal{E}}+1}^{3s_{e}+1,3s_{d}}\left[\frac{\mathcal{C}_{4}U_{d}}{\mathcal{B}_{4}U_{e}}\biggl|
\begin{array}{c}
1-l_{d_{2}},1,l_{e_{1}}\\
l_{e_{2}},0,1-l_{d_{1}}\\
\end{array}
\right].
\end{align}
%*************<END-EQUATION>************

\vspace{2mm}
%***************<HEADING>***************
\noindent
\textbf{Asymptotic Expression:}

Applying identical process as was performed for \eqref{a14} and \eqref{a20}, the asymptotic expression of IP in \eqref{a22} is expressed as
%***************<EQUATION>**************
\begin{align}
\label{a23}
IP_{\infty}={\sum_{q_{d,e}}}\sum_{p=1}^{3s_{d}}\frac{\prod_{l=1,l\neq p}^{3s_{d}}\Gamma(L_{1,p}-L_{1,l})\prod_{l=1}^{3s_{e}+1}\Gamma(1+L_{2,l}-L_{1,p})}{\prod_{l=3s_{d}+1}^{s_{\mathcal{M}}}\Gamma(1+L_{1,l}-L_{1,p})\prod_{l=3s_{e}+2}^{s_{\mathcal{E}}}\Gamma(L_{1,p}-L_{2,l})}\biggl(\frac{\mathcal{C}_{4}U_{d}}{\mathcal{B}_{4}U_{e}}\biggl)^{{L_{1,p}}-1}.
\end{align}
%*************<END-EQUATION>************

%%%%%%%%%%%%%%%%<SECTION>%%%%%%%%%%%%%%%
\section{Numerical Results}

In this section, we present the numerical results utilizing the deduced expressions of secrecy parameters i.e. SOP, SPSC, and IP. Besides, we also plot Monte-Carlo simulation results to validate our analysis via MATLAB. The AKM-shadowed and $\mathcal{M}$ random variables are generated via MATLAB, where we make an average of 100,000 channel realizations to obtain every value of secrecy parameters. The whole analysis is performed considering $\alpha\geq0$, $\kappa\geq0$, $\mu\geq0$, $\mathcal{T}_{SD}=1$, $\mathcal{T}_{c}=0.5$ bits/sec/Hz, $x\geq0$, $(a,b)$=($2.296,2$) for strong turbulence, ($4.2,3$) for moderate turbulence, and ($8,4$) for weak turbulence, $s_{d}=s_{e}=(1,2)$, and $\epsilon=1.1$ and $6.7$.
%===============<FIGURE>================
\begin{figure}[!ht]
\vspace{-25mm}
    \centerline{\includegraphics[width=0.6\textwidth,angle =0]{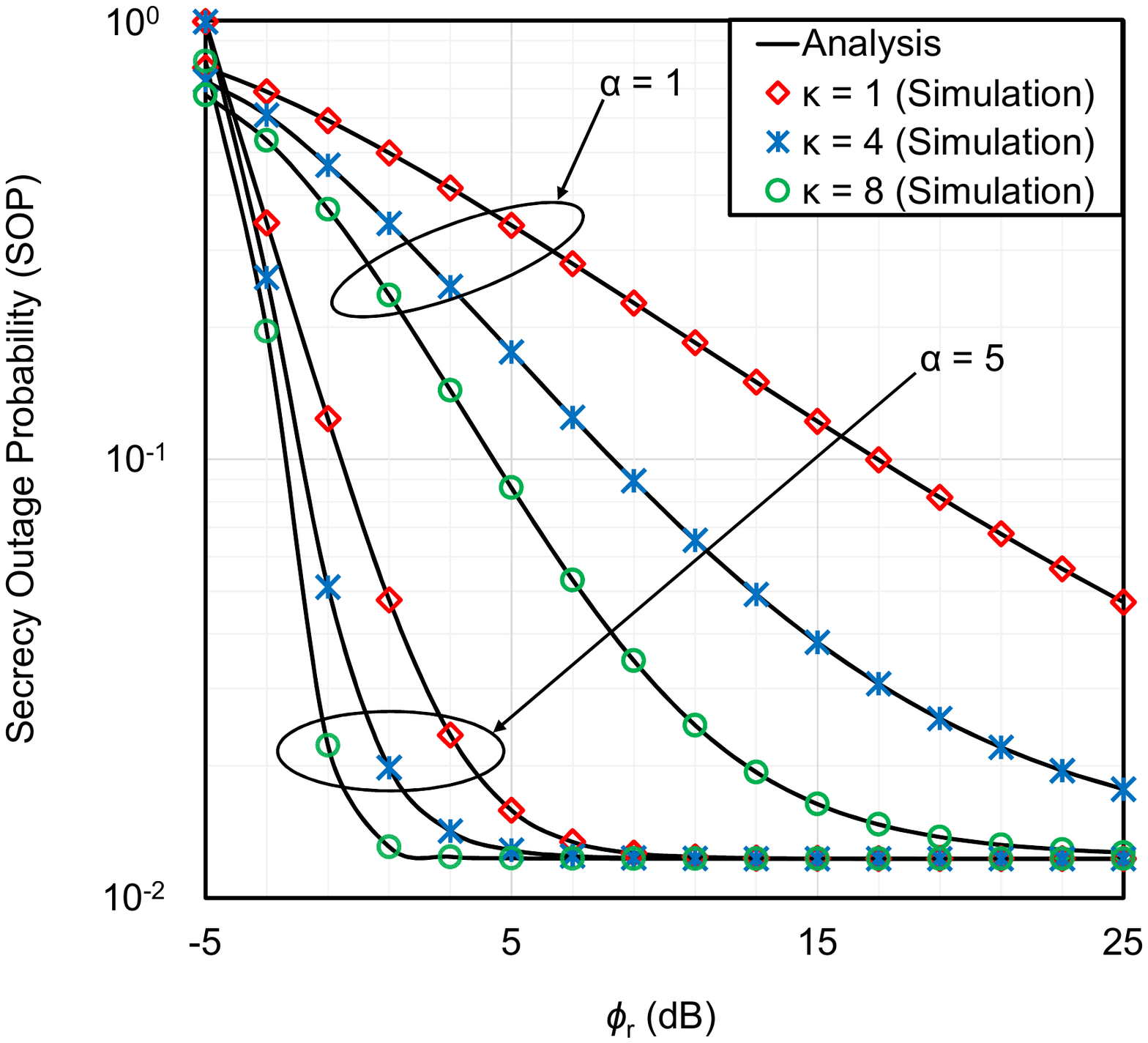}}
        \vspace{-30mm }
    \caption{
         SOP versus $\phi_{r}$ for selected values of $\alpha$ and $\kappa$ with $\mu=1$, $x=100$, $a=4.2$, $b=3$, $s_{d}=s_{e}=1$, $U_{d}=15dB$, $U_{e}=-5dB$, $\epsilon=1.1$, $r_{d}=r_{e}=0.1$, and $\zeta_{t_{d}}=\zeta_{t_{e}}=1$.
    }
    \label{g3}
\end{figure}
%=============<END-FIGURE>==============
%===============<FIGURE>================
\begin{figure}[!ht]
\vspace{-25mm}
    \centerline{\includegraphics[width=0.6\textwidth,angle =0]{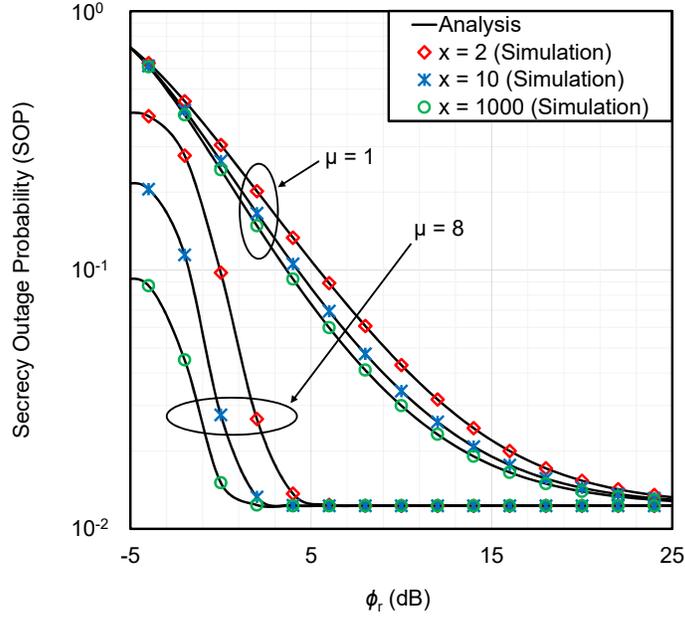}}
        \vspace{-30mm }
    \caption{
        SOP versus $\phi_{r}$ for selected values of $\mu$ and $x$ with $\alpha=\kappa=2$, $a=4.2$, $b=3$, $s_{d}=s_{e}=1$, $U_{d}=10dB$, $U_{e}=-10dB$, $\epsilon=1.1$, $r_{d}=r_{e}=0.1$, and $\zeta_{t_{d}}=\zeta_{t_{e}}=1$.
    }
    \label{g4}
\end{figure}
%=============<END-FIGURE>==============

Figures \ref{g3} and \ref{g4} depict the impact of channel parameters of AKM-shadowed fading channel (i.e. $x$, $\alpha$, $\kappa$, and $\mu$) on secrecy performance of the proposed system. For this purpose, the SOP is plotted against $\phi_{r}$ in both figures. It can clearly be seen that with the increase in $x$, $\alpha$, $\kappa$, and $\mu$, the SOP significantly decreases as testified in \cite{ramirez2019alpha}. In fact, an increase in $\alpha$, $\kappa$, and $\mu$ decreases the overall fading thereby improving the secrecy performance. Additionally, higher values of $x$ denotes lower amount of shadowing and hence the SOP decreases with $x$.
%===============<FIGURE>================
\begin{figure}[!ht]
\vspace{-25mm}
    \centerline{\includegraphics[width=0.6\textwidth,angle =0]{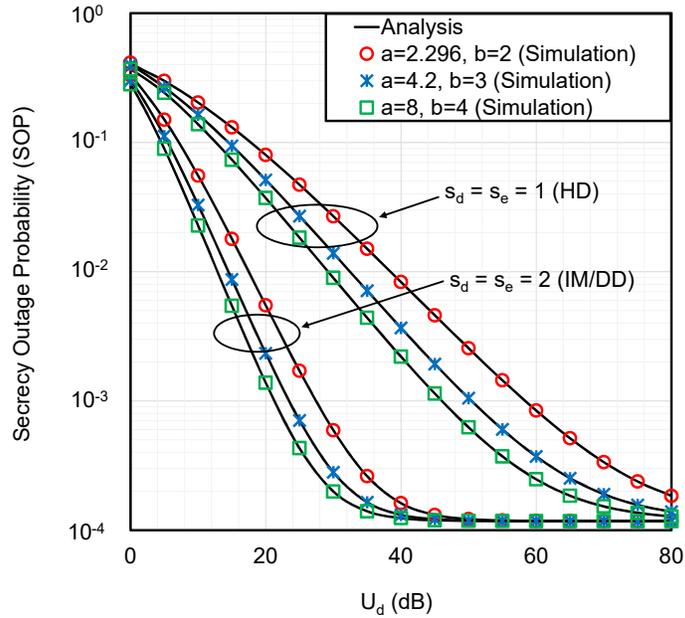}}
        \vspace{-30mm}
    \caption{
        SOP versus $U_{d}$ for selected values of $a$, $b$, $s_{d}$, and $s_{e}$ with $\alpha=2.5$, $\kappa=\mu=2$, $x=1000$, $\phi_{r}=10dB$, $U_{e}=-5dB$, $\epsilon=1.1$, $r_{d}=r_{e}=0.1$, and $\zeta_{t_{d}}=\zeta_{t_{e}}=1$.
    }
    \label{g5}
\end{figure}
%=============<END-FIGURE>==============
%===============<FIGURE>================
\begin{figure}[!ht]
\vspace{-10mm}
    \centerline{\includegraphics[width=0.75\textwidth,angle =0]{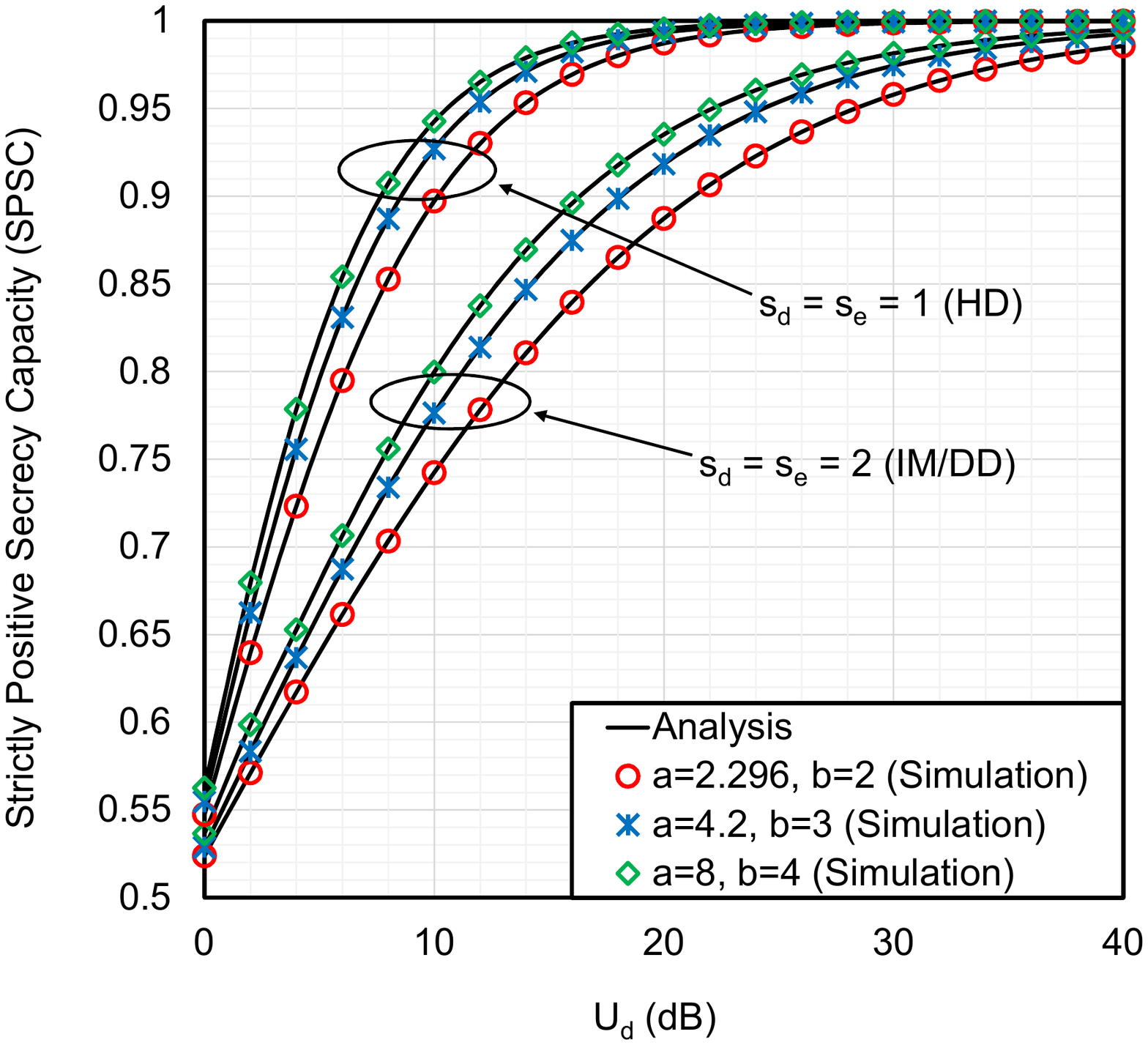}}
        \vspace{-10mm }
    \caption{
        SPSC versus $U_{d}$ for selected values of $a$, $b$, $s_{d}$, and $s_{e}$ with $U_{e}=-1dB$, $\epsilon=1.1$, $r_{d}=r_{e}=0.1$, and $\zeta_{t_{d}}=\zeta_{t_{e}}=1$.
    }
    \label{g6}
\end{figure}
%=============<END-FIGURE>==============
%===============<FIGURE>================
\begin{figure}[!ht]
\vspace{-25mm}
    \centerline{\includegraphics[width=0.6\textwidth,angle =0]{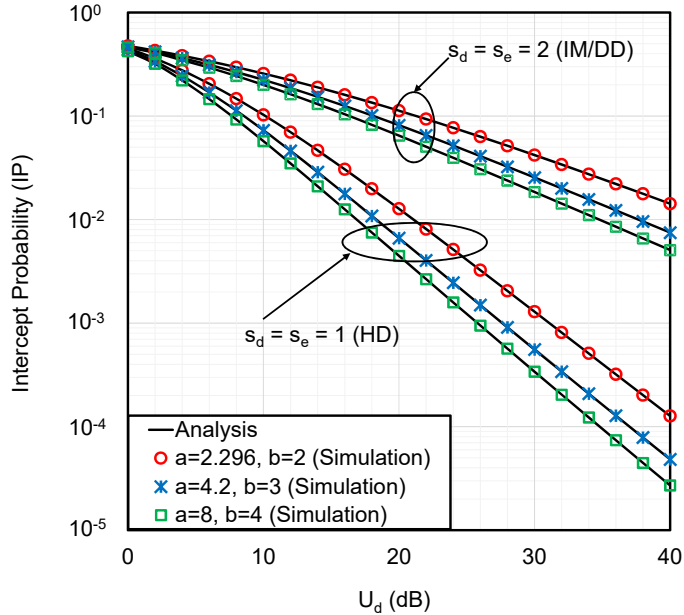}}
        \vspace{-30mm }
    \caption{
         IP versus $U_{d}$ for selected values of $a$, $b$, $s_{d}$, and $s_{e}$ with $U_{e}=-1dB$, $\epsilon=1.1$, $r_{d}=r_{e}=0.1$, and $\zeta_{t_{d}}=\zeta_{t_{e}}=1$.
    }
    \label{g7}
\end{figure}
%=============<END-FIGURE>==============

Impact of two types of detection techniques (HD and IM/DD) at the receiver and eavesdropper on secrecy performance are demonstrated in Figs. \ref{g5}-\ref{g7}. Results demonstrate that better secrecy performance can be achieved while employing HD technique ($s_{d}=s_{e}=1$) at both $D$ and $E$ relative to IM/DD technique ($s_{d}=s_{e}=2$). The reason behind this outcome is due to the fact of obtaining a better SNR at the destination with HD technique compared to IM/DD technique. Our exhibited results also match with the results exhibited in \cite{pan2019secrecy, pattanayak2020physical}.
%===============<FIGURE>================
\begin{figure}[!ht]
\vspace{-10mm}
    \centerline{\includegraphics[width=0.75\textwidth,angle =0]{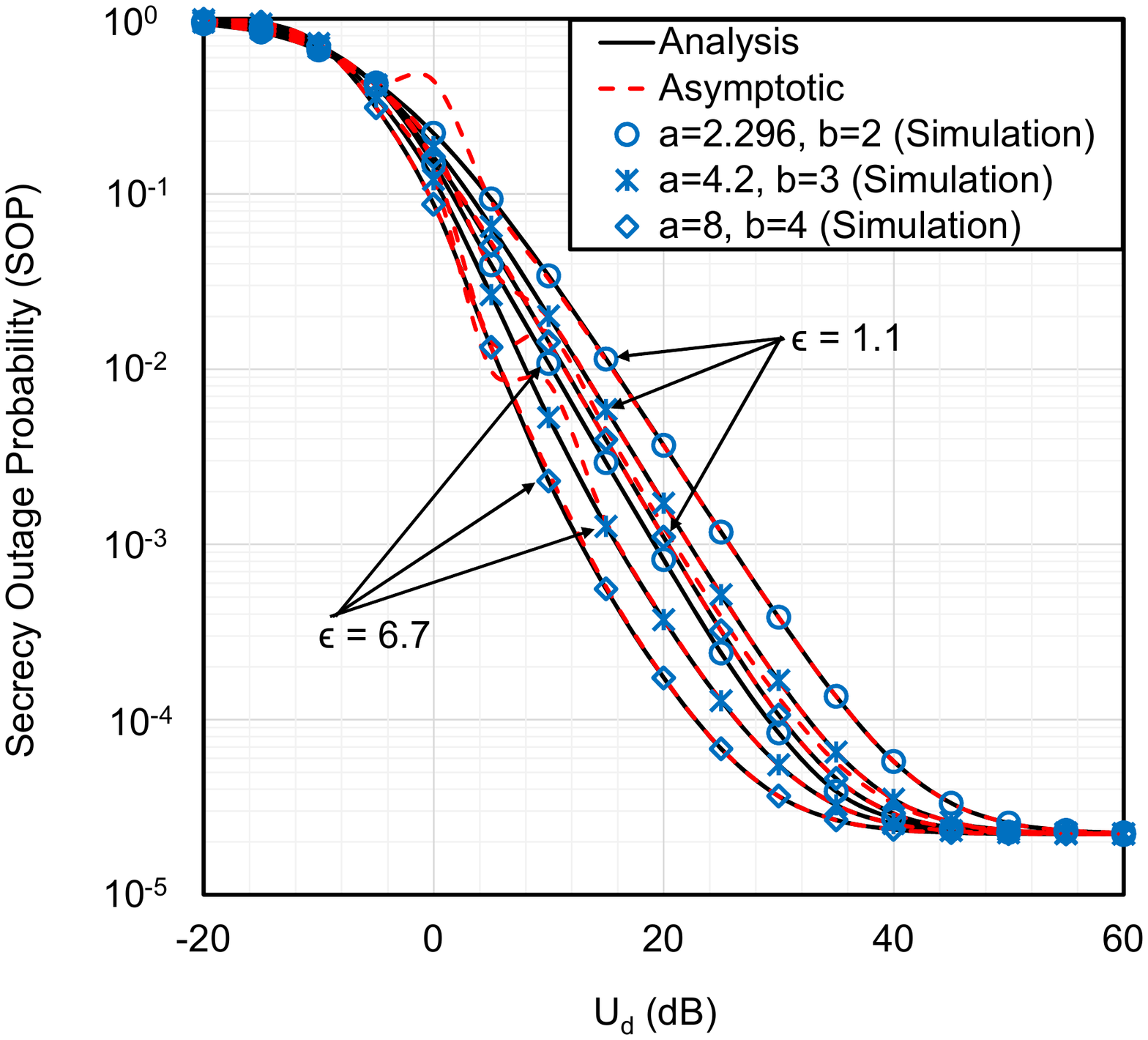}}
        \vspace{-10mm }
    \caption{
         SOP versus $U_{d}$ for selected values of $a$, $b$, and $\epsilon$ with $\alpha=3$, $\kappa=\mu=2$, $x=1000$, $\phi_{r}=12dB$, $s_{d}=s_{e}=1$, $U_{e}=-10dB$, $r_{d}=r_{e}=0.1$, and $\zeta_{t_{d}}=\zeta_{t_{e}}=1$.
    }
    \label{g8}
\end{figure}
%=============<END-FIGURE>==============
%===============<FIGURE>================
\begin{figure}[!ht]
\vspace{-10mm}
    \centerline{\includegraphics[width=0.75\textwidth,angle =0]{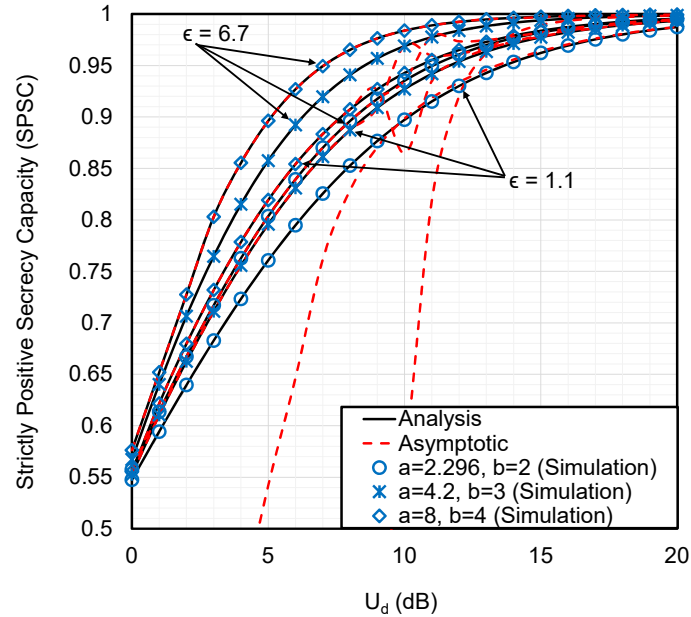}}
        \vspace{-10mm }
    \caption{
        SPSC versus $U_{d}$ for selected values of $a$, $b$, and $\epsilon$ with $s_{d}=s_{e}=1$, $U_{e}=2dB$, $r_{d}=r_{e}=0.1$, and $\zeta_{t_{d}}=\zeta_{t_{e}}=1$.
    }
    \label{g9}
\end{figure}
%=============<END-FIGURE>==============
%===============<FIGURE>================
\begin{figure}[!ht]
\vspace{-25mm}
    \centerline{\includegraphics[width=0.6\textwidth,angle =0]{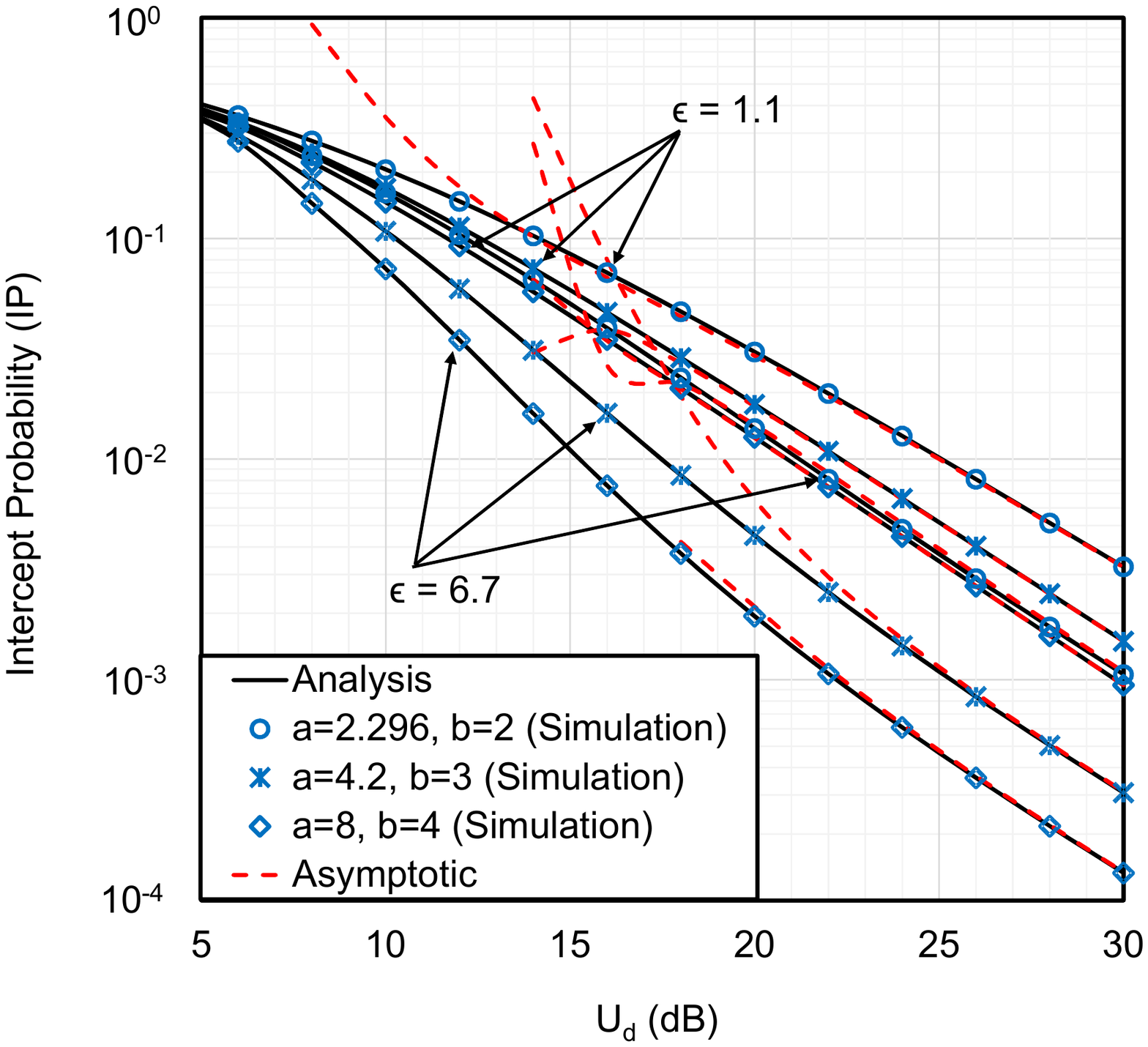}}
        \vspace{-30mm }
    \caption{
         IP versus $U_{d}$ for selected values of $a$, $b$, and $\epsilon$ with $s_{d}=s_{e}=1$, $U_{e}=3dB$, $r_{d}=r_{e}=0.1$, and $\zeta_{t_{d}}=\zeta_{t_{e}}=1$.
    }
    \label{g10}
\end{figure}
%=============<END-FIGURE>==============

The influence of pointing error on the secrecy performance is analyzed in Figs. \ref{g8}-\ref{g10} by depicting SOP, SPSC, and IP against $U_{d}$. Results reveal that when the FSO link experiences severe ($\epsilon=1.1$) to negligible pointing error ($\epsilon=6.7$), secrecy performance improves dramatically. This is because a lower pointing error indicates better pointing accuracy. Similar impacts of pointing error were also experienced in \cite{pan2019secrecy, pattanayak2020physical} that proves our analytical and simulation results are accurate. To gain further insights, we also provide asymptotic analysis and it is noteworthy that in a high SNR regime, the simulation, asymptotic, and analytical results match tightly with each other.

Besides the detection types and pointing error, atmospheric turbulence also affects secrecy performance as demonstrated in Figs. \ref{g5}-\ref{g10}. Analytical and simulation results indicate that similar to results in \cite{pan2019secrecy} and \cite{pattanayak2020physical}, our secrecy performance is the best at weaker turbulence scenarios and vice versa holds true too. These outcomes are obvious as severe turbulence affects the received SNR at the destination quite drastically relative to weaker turbulence.
%===============<FIGURE>================
\begin{figure}[!ht]
\vspace{-25mm}
    \centerline{\includegraphics[width=0.6\textwidth,angle =0]{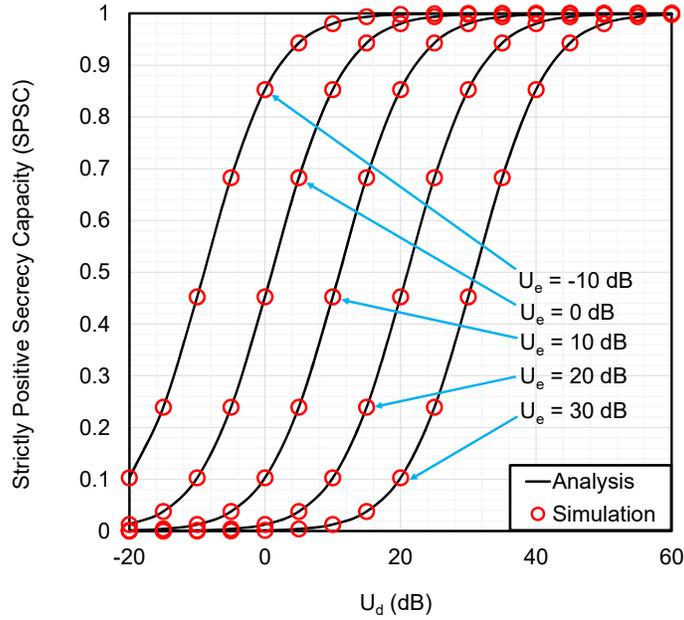}}
        \vspace{-30mm}
    \caption{
         SPSC versus $U_{d}$ for selected values of $U_{e}$ with $a=2.296$, $b=2$, $s_{d}=s_{e}=1$, $\epsilon=1.1$, $r_{d}=r_{e}=0.1$, and $\zeta_{t_{d}}=\zeta_{t_{e}}=1$.
    }
    \label{g11}
\end{figure}
%=============<END-FIGURE>==============
%===============<FIGURE>================
\begin{figure}[!ht]
\vspace{-25mm}
    \centerline{\includegraphics[width=0.6\textwidth,angle =0]{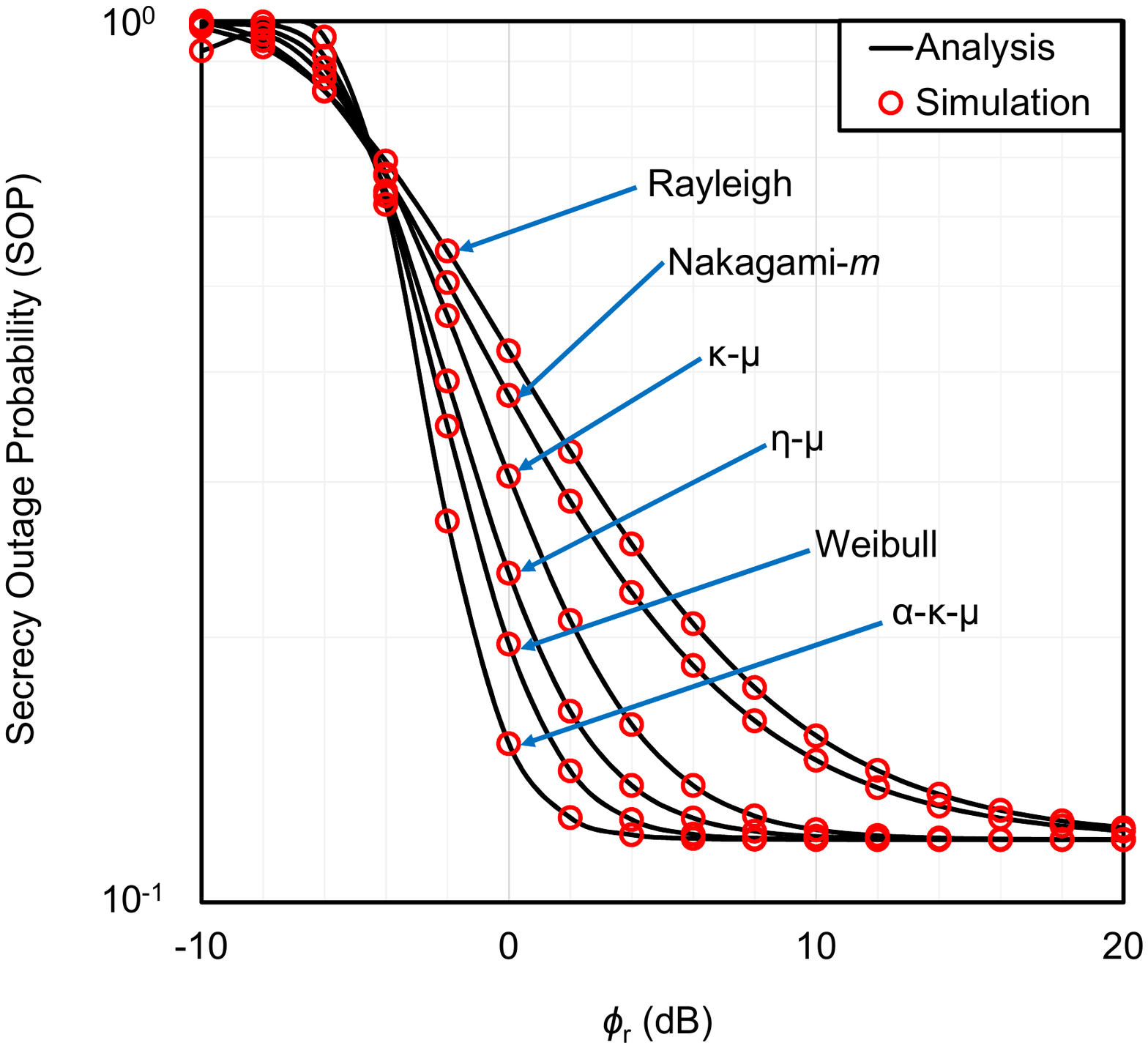}}
        \vspace{-30mm }
    \caption{
         SOP versus $\phi_{r}$ for selected values of $\alpha$, $\kappa$, $\mu$, and $x$ with $a=2.296$, $b=2$, $s_{d}=s_{e}=1$, $U_{d}=15dB$, $U_{e}=0dB$, $\epsilon=1.1$, $r_{d}=r_{e}=0.1$, and $\zeta_{t_{d}}=\zeta_{t_{e}}=1$.
    }
    \label{g1}
\end{figure}
%=============<END-FIGURE>==============
%===============<FIGURE>================
\begin{figure}[!ht]
\vspace{-25mm}
    \centerline{\includegraphics[width=0.6\textwidth,angle =0]{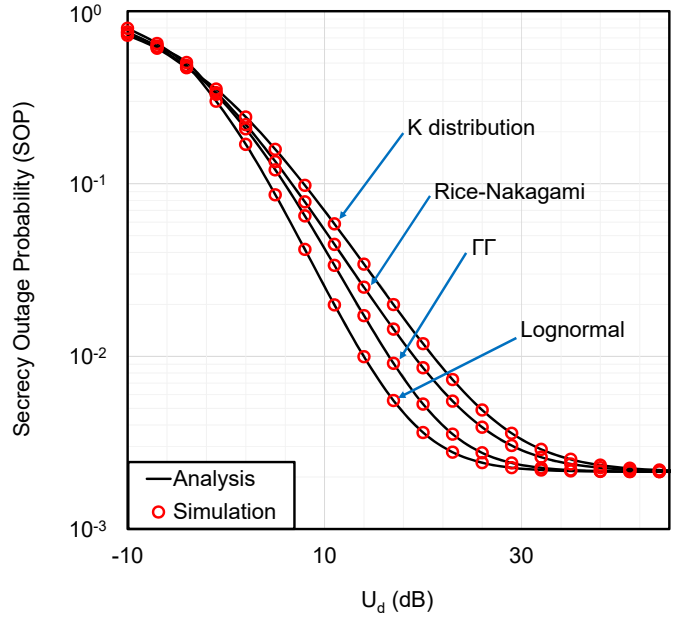}}
        \vspace{-30mm }
    \caption{
         SOP versus $U_{d}$ for selected values of $r_{d}$, $r_{e}$, $\zeta_{t_{d}}$, and $\zeta_{t_{e}}$ with $\alpha=3$, $\kappa=1$, $\mu=2$, $x=1000$, $\phi_{r}=5dB$, $a=4.2$, $b=3$, $s_{d}=s_{e}=1$, $U_{e}=-5dB$, and $\epsilon=1.1$.
    }
    \label{g2}
\end{figure}
%=============<END-FIGURE>==============

We observe the impact of electrical SNR $U_{e}$ of the $R-E$ link in Fig. \ref{g11} with respect to SPSC.
Our results demonstrate the expected outcome as SPSC decreases when $U_{e}$ increases from a lower to a higher value. This occurs since higher $U_{e}$ signifies a stronger $R-E$ link. A similar type of result was also exhibited in \cite{pattanayak2020physical} that strongly justifies our results.

The generic characteristics of AKM-shadowed fading is demonstrated in Fig. \ref{g1} following Table \ref{t1}. It is observed from Fig. \ref{g1} that not only multipath fading channels (e.g. Rayleigh, Nakagami-$m$, and Weibull), but also generalized fading channels (e.g. $\eta-\mu$, $\kappa-\mu$, and $\alpha-\kappa-\mu$) can be obtained as special cases to our proposed RF model. Figure \ref{g2} demonstrates the generic characteristics of $\mathcal{M}$ distribution by utilizing the parameter values as presented in Table \ref{t2}.

It can clearly be observed that $K$ distribution, Rice-Nakagami, $\Gamma\Gamma$, Lognormal, etc., models can easily be obtained as special cases to our work.

\quad

\noindent
\textbf{Comparative Analysis with Existing Related Literature:}
%===============<FIGURE>================
\begin{figure}[!ht]
\vspace{-10mm}
    \centerline{\includegraphics[width=0.75\textwidth,angle =0]{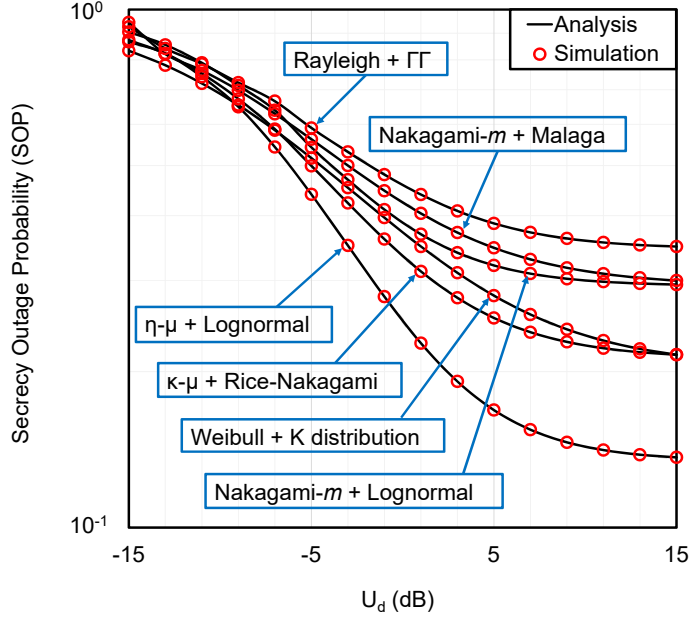}}
        \vspace{-10mm }
    \caption{
         SOP versus $U_{d}$ for selected values of $\alpha$, $\kappa$, $\mu$, $x$, $r_{d}$, $r_{e}$, $\zeta_{t_{d}}$, $\zeta_{t_{e}}$, and $b$ with $\phi_{r}=0dB$, $a=4.2$, $s_{d}=s_{e}=1$, $U_{e}=-10dB$, and $\epsilon=1.1$.
    }
    \label{g12}
\end{figure}
%=============<END-FIGURE>==============

We assume our RF link experiences the AKM-shadowed fading model whereas the FSO link experiences the unified $\mathcal{M}$ turbulence with pointing error impairment. AKM-shadowed is a composite fading model that comprises a large number of multipath and generalized fading models as listed in Table \ref{t1}. On the other hand, the $\mathcal{M}$ turbulence model also houses immense generic characteristics (Table \ref{t2}) that make it one of the most popular FSO turbulence models among optical wireless communication researchers. Hence, our hybrid RF-FSO system model can unify a wide range of both existing and non-existing RF-FSO hybrid scenarios for which a summary is deduced in Table \ref{t3}. Subsequently, Fig. \ref{g12} demonstrates this generalization graphically wherein we can clearly observe the proposed model exhibits significant generality and novelty relative to the open literature.
%<<<<<<<<<<<<<<<<<TABLE>>>>>>>>>>>>>>>>>
\begin{table*}[!ht]
\centering
\caption{Special Cases of Our Proposed Model}
\scalebox{0.90}{%
\begin{tabular}{|>{\centering\arraybackslash}p{2.5 cm}|>{\centering\arraybackslash} p{7cm}|>{\centering\arraybackslash}p{8cm}|}
\hline
Reference Model & RF link & FSO link
\\ 
\hline
\hline
- & Nakagami-$m$ ($\alpha=2$, $\kappa=0$, $\mu=x=2$) & Lognormal ($\zeta_{t_{m}}=2$, $r_{m}=0.0001$, $b=3$)
\\
\hline
- & Weibull ($\alpha=3$, $\kappa=0$, $\mu=x=1$) & K distribution ($\zeta_{t_{m}}=2$, $r_{m}=0.1$, $b=1$)
\\
\hline
- & $\eta-\mu$ ($\alpha=2$, $\kappa=0$, $\mu=4$, $x=2$) & Log-normal ($\zeta_{t_{m}}=2$, $r_{m}=0.0001$, $b=3$)
\\
\hline
- & $\kappa-\mu$ ($\alpha=2$, $\kappa=1$, $\mu=2$, $x=100$) & Rice-Nakagami ($\zeta_{t_{m}}=2$, $r_{m}=0.1$, $b=3$)
\\
\hline
\cite{pan2019secrecy} & Rayleigh ($\alpha=2$, $\kappa=0$, $\mu=x=1$) & $\Gamma\Gamma$ ($\zeta_{t_{m}}=1$, $r_{m}=0$, $b=2$)
\\
\hline 
\cite[(Scenario-2)]{pattanayak2020physical} & Nakagami-$m$ ($\alpha=2$, $\kappa=0$, $\mu=x=2$) & M\'alaga ($\zeta_{t_{m}}=1$, $r_{m}=0.1$, $b=3$)
\\
\hline
\end{tabular}}
\label{t3}
\end{table*}
%<<<<<<<<<<<<<<<<END-TABLE>>>>>>>>>>>>>>

%%%%%%%%%%%%%%%%<SECTION>%%%%%%%%%%%%%%%
\section{Conclusions}

This work focuses on the protection of secret information against FSO eavesdropping over a RF-FSO mixed system where the RF and FSO links are assumed to follow AKM-shadowed and M\'alaga turbulence fading models. Secrecy analysis was carried out in terms of closed-form expressions for three secrecy metrics i.e. SOP, SPSC, and IP that were validated via Monte-Carlo simulations. Additionally, we also investigated the asymptotic expressions for each metric to demonstrate more useful insights and tractability, and it is observed the asymptotic expressions exhibit appropriate tightness in high SNR regimes. Numerical results reveal that fading, shadowing, atmospheric turbulence, and pointing misalignment error have tremendous detrimental impacts on the secrecy performance. Moreover, the HD technique always outperforms IM/DD technique. Since, in our considered scenario, both RF and FSO links are generalized, our demonstrated results exhibit superiority over the existing literature via providing these results to the design engineers while working on more real-life systems considering higher order of randomness in the propagation channel.
%%%%%%%%%%%%%%%%%%%%%%%%%%%%%%%%%%%%%%%%%%%%%%%
\bibliography{main.bib}

\end{document}